\documentclass[10pt,a4paper,twoside,twocolumn,nofootinbib]{revtex4}

\usepackage{amsfonts,amssymb,amsmath}     
\usepackage{graphicx,pst-all,epsfig}      
\usepackage[latin1]{inputenc}             
\usepackage{latexsym}                     
\usepackage{mathrsfs,bm}                  
\usepackage{hyperref,float}               
\usepackage{afterpage}                    
\usepackage{longtable}                    
\usepackage{fancyhdr}                     
\usepackage{lastpage}                     
\usepackage{pgf,setspace}

\begin{document}


\title{Casimir energy for the scalar field: a global approach with cut-off exponential function}


\author{M.S.R. Miltão}
\email[Endere\c{c}o Eletrônico: ]{miltaaao@ig.com.br}

\author{Franz A. Farias}
\email[Endere\c{c}o Eletrônico: ]{franz\_farias@hotmail.com}

\affiliation{Departamento de F\'{\i}sica -- UEFS \\ Av. Transnordestina, Km 03, BR 116 \\ Feira de Santana -- BA -- Brazil -- 44036-900}


\begin{abstract}
A global approach with cut-off exponential functions previously proposed is used to obtain the Casimir energy of a massless scalar field in the presence of a spherical shell. This method makes the use of two regulators, one of them to turn the sum of the orders of Bessel functions finite and a second, to turn the integral involving the zeros of Bessel function regularized. This proposed procedure ensures a consistent mathematical handling in the calculations of the Casimir energy for a scalar field as well it does show all types of divergences of interest. We separately consider the contributions of the inner and outer regions of a spherical shell and we show that the results obtained are in agreement with those known in the literature and this gives a confirmation for the consistence of the proposed approach.
\medskip

\end{abstract}

\maketitle


\section{Introduction}

\indent

The relevance of the Casimir effect has increased over the decades since the seminal paper (1948) \cite{Casimir48} by the Dutch Physicist Hendrik Casimir. This effect concerns to the appearance of an attractive force between two plates when they are placed close to each other. H. Casimir was the first to predicted and explains the effect as a change in vacuum quantum fluctuations of the electromagnetic field.

Nowadays, the Casimir effect has been applied to a variety of quantum fields and geometries and it has gained a wider understanding as the effect which comes from the fluctuations of the zero point energy of a relativistic quantum field due to changes in its base manifold. This interpretation can be confirmed when we see the large range where the Casimir effect has been applied: the study of gauge fields with BRS symmetry \cite{Ambjorn83}, in the Higgs fields \cite{Aoyama84}, in supersymmetric fields \cite{Igarashi84}, in supergravity theory \cite{Inami84}, in superstrings \cite{Kikkawa84}, in the Maxwell-Chern-Simons fields \cite{Milton90}, in relativistic strings \cite{Brevik97}, in M-theory \cite{Fabinger00}, in cosmology \cite{Zerbini01}, in non-commutative spacetimes \cite{Demetrian02}, among others subjects in the literature \cite{Maclay01,Miltao04}, the review articles \cite{Greiner86,Lamoreaux99,Bordag01} and textbooks \cite{Milonni94,Elizalde95,Mostepanenko97,Milton01}.

In this present work the meaning of base manifold is that the confinement that the field is subjected it is due to the presence of a sphere, where the boundary conditions takes place. The point we aim to emphasize is that once the calculation of the Casimir effect involves dealing with infinite quantities we need to use a regularization procedure appropriately defined. Many different regularization methods has been proposed and we can quote some of them: the summation mode method -- using the general cut-off function \cite{Casimir48}, or exponential cut-off function \cite{Davies72}, or Green function \cite{Lukosz73,Bender76,Sen81,Igarashi83}, or Green function through multiple scattering \cite{Balian78}, or exponential function and cut-off parameter \cite{Olaussen81}, or zeta function \cite{Ruggiero77,Hawking77,Bordag96,Romeo96,Vassilevich95,Bender96,Svaiter01}, or Abel-Plana formula \cite{Mostepanenko97,Saharian00}, or point-splitting \cite{Fulling76}; the Green function method -- using the point-splitting \cite{Brown69,Christensen76,Candelas79}, or Schwinger's source theory \cite{Schwinger78}, or zeta function \cite{Dowker76}; and the statistical approach method -- using the path integral formalism \cite{Hoye01}, or Green function \cite{Klich01}; as some examples among others. These methods are distinguished by the approach used to carry out the calculations of the Casimir energy and it is clear that the physical result must be independent from the regulators or the method employed for them. But the literature has shown that the results found there, exhibit a divergence among them.

In a general way, the methods used to obtain the Casimir effect lies on one of the two categories: a local procedure or a global one. With a local procedure we mean one that the expression for the change of the vacuum energy is explicitly dependent on the variables of the base manifold and only in the final step of calculations the integration over these variables is carried out. On the other hand, in a global one we start with an expression for the vacuum energy where there is no space-time variables present as they already were integrated.

In previous works \cite{Miltao06,Miltao08} it was established a global approach for the calculation of the spherical electromagnetic Casimir effect. There it was proposed the use of two regulators into the cut-off exponential function and it was demonstrated that this regularization approach is one appropriate for the calculation of Casimir effect in the case of a spherical symmetry. Now we apply the method to the situation when we have a massless scalar field in the presence of a spherical shell. The interest lies in presenting the consistence of the proposed method when we compare the obtained results with those in the literature which are acquire by other methods. With the scalar field we can avoid the inherent complications brought by the vector nature of the electromagnetic field and due to its simple structure the scalar field usually has become an effective tool to be used in the investigation of field proprieties as examples: in the dynamical Casimir effect \cite{Moussa08}, in the Casimir effect at finite temperature \cite{Setare03,Teo09,Ttira08}, in the Casimir effect on a presence of a gravitational field \cite{Setare05,Napolitano08}, among others \cite{Moylan08,Pedraza08,Jaffe08}.

The paper is organized as follows: we detail in the section 2 the method to be used and why we need two cut-off parameters to obtain an regularized expression for the Casimir energy, which it is the starting point for a consistent mathematical handling. The section 3 exhibits the calculations for the contributions of the inner and the outer regions of the spherical shell. We analyze in the section 4 the results and compare them with those ones in the literature and make some considerations.


\section{The global procedure proposed with two parameters}

\indent

The starting point is the expression for the Casimir energy defined as the difference between the vacuum energy under a give boundary condition and the reference vacuum energy. When we consider a scalar field in the presence of a spherical shell this vacuum energy is
\begin{eqnarray}
E _{0} = \sum _{n = 1} ^{\infty} \sum _{j = 0} ^{\infty} \sum _{m = -j} ^{j} \sum _{\tau} \frac{1}{2} \hbar \omega _{j n } ^{\tau}, \label{01}
\end{eqnarray}
where $\omega _{j n } ^{\tau} $ are the mode frequencies. They are obtained when the boundary conditions are imposed on the field. In the absence of boundary conditions the frequencies take some values which let us designate as $\omega _{j n } ^{\tau (ref)} $ and these lead to the vacuum reference energy
\begin{eqnarray}
E ^{(ref)} = \sum _{n = 1} ^{\infty} \sum _{j = 0} ^{\infty} \sum _{m = -j} ^{j} \sum _{\tau} \frac{1}{2} \hbar \omega _{j n } ^{\tau (ref)}, \label{02}
\end{eqnarray}
so, the Casimir energy is $\mathcal{E} = E _{0} - E ^{(ref)} $. The boundary conditions due to a spherical shell with radius $a$ are
\begin{eqnarray}
kaj_{j}\left( ka\right) &=&0 , {\text{ for }} r=a-0 , \text{ \ } \label{04} \\
A _{j} k a j _{j} \left( k a \right) + B_{j}kan_{j}\left( k a \right) &=& 0, {\text{ for }} r=a+0.  \text{ \ } \label{05}
\end{eqnarray}

\begin{figure}[H]
\begin{center}
  \includegraphics[width=6cm]{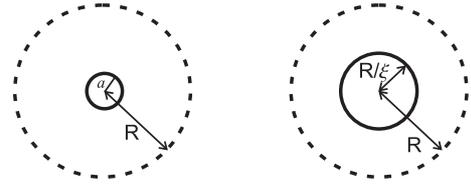}
  \caption{A sketch of the subtraction process which takes place on the regularization of the Casimir effect of a spherical shell.} \label{SubCasca}
\end{center}
\end{figure}

The Casimir energy will be calculated by using the mode summation and the argument theorem (also known as argument principle \cite{Ahlfors79,LinsNeto96,Spiegel97}). This theorem gives the summation of zeros and poles of an analytic function as a contour integral. This contour is a curve that encompass the interior region of the complex plane which contains the zeros and poles \cite{Ahlfors79,LinsNeto96,Spiegel97}. In our case we are interested in the root functions which match the conditions (\ref{04}) and (\ref{05}). So, the following are appropriate as root functions
\begin{eqnarray}
f_{j}^{1}\left( az\right) &=& azj_{j}\left( az\right) ,  \label{07} \\
f_{j}^{2}\left( az\right) &=&\cos \delta _{j}\left( z\right) \left[ azj_{j}\left( az\right) + \tan \delta _{j}\left( z\right) azn_{j}\left( az\right) \right] , \text{ \ \ \ } \label{08}
\end{eqnarray}
where
\begin{eqnarray}
z = k \left( =\frac{\omega }{c}\right) , \text{ \ and \ } \delta _{j}\left( z\right) = z R-\frac{j\pi }{2} . \label{10}
\end{eqnarray}
When we apply the argument theorem and carry out some handling, we get
\begin{equation}
\sum_{n=1}^{g}\omega_{jn}^{\tau} =\frac{c}{2\pi i}\oint_{C} dz \, z\frac{d}{dz} \log{f_{j} ^{\,\tau}(az)} . \label{11}
\end{equation}
On the above equation, the argument for logarithm must involves the product of all root functions. The contour to be taken on the calculations is given by \cite{Bowers98}
\begin{figure}[H]
\begin{center}
  \includegraphics[width=4cm]{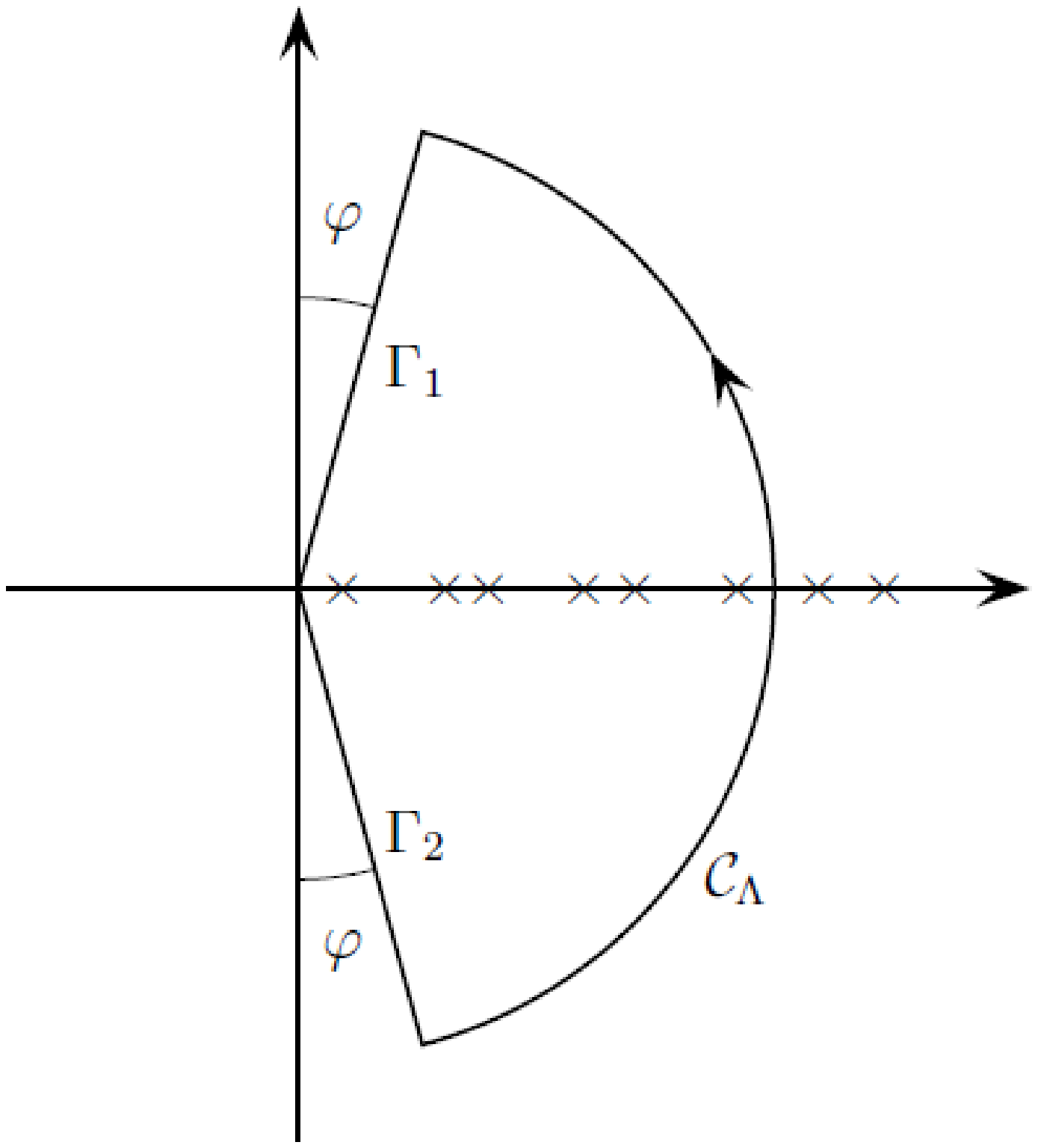}
  \caption{The path of integration in the complex plane.} \label{Percurso}
\end{center}
\end{figure}
The subtraction process (subtraction), defined by $\mathcal{E} $, can be schematically represented as in figure (\ref{SubCasca})

The vacuum energy (\ref{01}) which takes into account the boundary conditions can be used to obtain the reference energy in (\ref{02}). This is done when we take the limit for the radius $a$ going to infinity. This procedure is sensible but it already has been made clear by Boyer \cite{Boyer70}. After all, we obtain for the Casimir energy
\begin{eqnarray}
\mathcal{E}&=&\sum_{n=1}^{\infty }\sum_{j=0}^{\infty }\sum_{m=-j}^{j}\sum_{\tau =1} ^{4} \frac{1}{2} \hbar (\omega _{jn}^{\tau }-\omega _{jn}^{\tau (ref)}) \nonumber \\
&=& \lim_{\substack{ _{\substack{ \sigma \rightarrow 0  \\ \epsilon \rightarrow 0}}  \\ R\rightarrow \infty \\ _{\xi\rightarrow 1} }}\frac{\hbar c}{2\pi i} \sum_{j=0}^{\infty }\nu \exp \left( -\epsilon \nu \right) \oint_{C}dzz\exp \left( -\sigma z\right)\times   \nonumber \\
&\times&\frac{d}{dz} \left\{\log \left[f_{j}^{(1)}\left( az\right)f_{j}^{(2)}\left( az \right)\right]- \log\left[f_{j}^{(1)(ref)}\left(\frac{R}{\xi} z\right) \right.\right. \nonumber \\ &\times& \left. \left. f_{j}^{(2)(ref)} \left(\frac{R}{ \xi} z\right) \right] \right\},  \label{12}
\end{eqnarray}
where $\nu =j+1/2$. We can see from above that two exponential functions were used, one of them is the function under the integral sign, $\exp{(-\sigma \,z)}$, ($\sigma>0$), that stems from the argument theorem and the other the function $\exp{(-\epsilon \,\nu)}$, ($\epsilon>0$), under summation sign on $j=\nu-(1/2)$. Now group together these two developments, and the equation (\ref{12}) may be rewritten as
\begin{eqnarray}
\mathcal{E}&=&-\frac{\hbar c}{\pi}{\Re }\sum_{j=0}^{\infty }\nu ^{2}\exp (-\epsilon \nu )\int_{0}^{\infty \exp\left( -i\varphi \right) } dz\exp \left( -i\sigma \nu z\right) z  \nonumber \\ &\times &\frac{d}{dz} \left\{ \log \left[ I_{\nu }\left( \nu az\right) \right] + \log \left[ K _{\nu }\left( \nu az\right) \right] \right\} - E ^{(ref)}, \label{13}
\end{eqnarray}
where the limits for $R$ and $\xi$, has been taking into account. The other limits will be take in an appropriate moment after the cancelation of possible remained divergences.


\section{Casimir effect of a spherical shell -- The case of a scalar field}

\indent

We now rewrite the Eq. (\ref{13}) in a more appropriate way so that the contributions can be  separated by regions as $\mathcal{E} = \mathcal{E} _{I} +\mathcal{E} _{O} $,
where
\begin{eqnarray}
\mathcal{E} _{I} &=& -\frac{\hbar c}{\pi}{\Re } \left( \frac{1}{2} \right) ^{2} \exp (- \epsilon \frac{1}{2} )\int_{0}^{\infty \exp\left( -i\varphi \right) } dz z  \nonumber \\ &\times& \exp \left( -i\sigma \frac{1}{2} z\right)\frac{d}{dz} \left\{ \log \left[ I_{\frac{1}{2} }\left( \frac{1}{2} az\right) \right] \right\} \nonumber \\ &-&\frac{\hbar c}{\pi}{\Re } \sum _{j=1} ^{\infty }\nu ^{2}\exp (-\epsilon \nu )\int_{0}^{\infty \exp\left( -i\varphi \right) } dz z \nonumber \\ &\times & \exp \left( -i\sigma \nu z\right) \frac{d}{dz} \left\{ \log \left[ I_{\nu } \left( \nu az\right) \right] \right\} - E _{I} ^{ref}, \,\,\,\,\, \label{14a}
\end{eqnarray}
is the contribution due to the internal modes and
\begin{eqnarray}
\mathcal{E} _{O} &=& -\frac{\hbar c}{\pi}{\Re } \left( \frac{1}{2} \right) ^{2} \exp (- \epsilon \frac{1}{2} ) \int_{0}^{\infty \exp\left( -i\varphi \right) } dz z \nonumber \\ &\times& \exp \left( -i\sigma \frac{1}{2} z\right) \frac{d}{dz} \left\{ \log \left[ K _{ \frac{1}{2} } \left( \frac{1}{2} az \right) \right] \right\} \nonumber \\ &-&\frac{\hbar c}{\pi}{\Re }\sum_{j=1}^{\infty }\nu ^{2}\exp (-\epsilon \nu )\int_{0}^{\infty \exp\left( -i\varphi \right) } dz z \nonumber \\ &\times& \exp \left( -i\sigma \nu z\right) \frac{d}{dz} \left\{ \log \left[ K _{\nu }\left( \nu az\right) \right] \right\} - E _{O} ^{ref}, \,\,\,\,\, \label{14b}
\end{eqnarray}
is the contribution due to the external modes. As it can be observed the above contributions were written in such a way that the term for $j = 0 $ was detached from the summation on $j $. This has been done to the effect of making explicit the term which we will focus attention as well as to take into account some developments already accomplished \cite{Miltao06}.

\medskip


\subsection{Internal mode}

\indent

Now we proceed with the calculations of Eq. (\ref{14a}) and the first step is to take the Debye expansion for the Bessel functions up to order $\mathcal{O} \left( \nu ^{-4} \right) $ \cite{Watson66,Stegun72}. 
So we have
\begin{eqnarray}
\mathcal{E} _{I} = E _{I} - E _{I} ^{(ref)}, \label{reg-inner}
\end{eqnarray}
where $E _{I} = E _{I,0} + E _{I,1} + E _{I,2} + E _{I,3} + E _{I,4} $ and the terms $E _{I,n} $ are given by
\begin{widetext}
\begin{eqnarray}
E _{I,0} &=& - \frac{\hbar c}{2 \pi}\Re \exp (-\epsilon \frac{1}{2} ) \exp (-i\varphi ) \int_{0}^{\infty } d\rho \exp (-i\sigma \rho \exp (-i \varphi) ) \left[ -\frac{1}{2} + \exp (-i \varphi)  a \rho \coth ( a \rho \exp (-i \varphi) ) \right], \label{17a} \\
E _{I,1} &=& \frac{\hbar c}{\pi a}\sum_{j=1}^{\infty }\nu ^{2}\int_{0}^{\infty}d\rho \left\{ \log\left[\mathcal{I} _{I} \left(\nu ,\rho\right)\right] -\sum_{k=1}^{4} \frac{\mathcal{U} _{(I,k)} \left(t\right)}{\nu ^k} \right\} \; ,\label{17b} \\
E _{I,2} &=& - \frac{\hbar c}{\pi}\Re \sum_{k=1} ^{4}\sum_{j=1}^{\infty }\nu ^{2-k}\exp (-\epsilon \nu ) \int_{0}^{\infty \exp (-i\varphi)} dz\exp (-i\sigma \nu z) z \frac{d}{dz}\mathcal{U}_{(I,k)} \left(t\right) \; ,\label{17c}\\
E _{I,3} &=&  \frac{\hbar c}{2 \pi}\Re \sum_{j=1}^{\infty }\nu ^{2} \exp (-\epsilon \nu ) \int_{0}^{\infty \exp (-i\varphi )}dz\exp (-i\sigma \nu z) \frac{a^{2}z^{2}}{1+a^{2}z^{2}} \; ,\label{17d}\\
E _{I,4} &=& - \frac{\hbar c}{\pi}\Re \sum_{j=1}^{\infty }\nu ^{3} \exp (-\epsilon \nu ) \int_{0}^{\infty \exp (-i\varphi )} dz\exp (-i\sigma \nu z) \sqrt{1+a^{2}z^{2}} \; ,\label{17e}
\end{eqnarray}
\end{widetext}
with the definitions \cite{Romeo96}
\begin{eqnarray}
\mathcal{I} _{I}\left(\nu ,\rho\right)&=& \sqrt{2 \pi \nu}\left(1+ \rho^{2}\right)^{\frac{1}{4}} \exp\left(-\nu \eta \right) I_{\nu}\left(\nu \rho\right), \label{18a} \\
\mathcal{U}_{(I,1)}\left(t\right)&=& \frac{t}{8}- \frac{5 t^{3}}{24}, \, \, \label{18b}
\end{eqnarray}
\begin{eqnarray}
\mathcal{U}_{(I,2)}\left(t\right)&=& \frac{t^{2}}{16}- \frac{3 t^{4}}{8} +\frac{5 t^{6}}{16}, \label{18c} \\
\mathcal{U}_{(I,3)}\left(t\right)&=& \frac{25t^{3}}{384}- \frac{531 t^{5}}{640}+\frac{221 t^{7}}{128}- \frac{1105 t^{9}}{1152}, \label{18d}
\end{eqnarray}
and
\begin{eqnarray}
\mathcal{U}_{(I,4)}\left(t\right)&=& \frac{13t^{4}}{128}- \frac{71 t^{6}}{32} +\frac{531 t^{8}}{64}- \frac{339 t^{10}}{32} + \frac{565 t^{12}}{128}, \nonumber \\ \label{18e}
\end{eqnarray}
The contributions (\ref{17d}) and (\ref{17e}) compound the zero order terms of the Debye expansion. The contribution (\ref{17b}) stems from small values of the angular momentum $j$ and its value already was determined by \cite{Romeo96}
\begin{eqnarray}
E _{I,1} = 0.00024 \frac{\hbar c}{\pi a} . \label{19}
\end{eqnarray}
The contributions (\ref{17c}--\ref{17e}) are calculated taking into account the Euler-Maclaurin formula with remainder \cite{Knopp64,Barton81} and these was calculated by \cite{Miltao06}
\begin{eqnarray}
E _{I,2} &=& \frac{\hbar c}{\pi}\Re \left\{\frac{8099}{63839} \frac{1}{a}+ \frac{7}{24}\frac{a}{\sigma^{2}} + \frac{11 }{192}\frac{1}{a}\log\left( \frac{\sigma}{a}\right)+ \right.\nonumber \\
&&\left.\frac{229}{40320}\frac{1}{a}\log\left(\epsilon\right) + i \frac{7801}{86684}\frac{1}{a} \right\}, \label{20a}\\
E _{I,3} &=& \frac{\hbar c}{\pi}\Re \left\{\frac{52529}{267528} \frac{1}{a} + i\left[ -\frac{1}{3}\frac{1}{\sigma}-\frac{3}{4}\frac{1}{\sigma \epsilon} - \frac{1}{2}\frac{1}{\sigma \epsilon^{2}} \right] \right\}, \nonumber \\ \label{20b}\\
E _{I,4} &=& \frac{\hbar c}{\pi}\Re \left\{\frac{7375}{85696} \frac{1}{a}- \frac{11}{24}\frac{a}{\sigma^{2}}+ 2 \frac{a ^{3}}{\sigma^{4}}-\frac{127}{1920} \frac{1}{a}\log\left(\frac{\sigma}{a}\right)\right.\nonumber\\
&&\left.-i\frac{2197}{21145}\frac{1}{a}\right\}. \label{20c}
\end{eqnarray}
Collecting the terms (\ref{19}), (\ref{20a}), (\ref{20b}) and (\ref{20c}) we get
\begin{eqnarray}
E _{I_{partial}} &=& 0.4095155894 \frac{\hbar c}{\pi a} + \frac{\hbar c}{\pi} \left[-\frac{1}{6}\frac{a}{\sigma^{2}} + 2 \frac{a^{3}}{\sigma^{4}}  \right.\nonumber\\
&-&\left.\frac{17}{1920} \frac{1}{a} \log\left( \frac{\sigma}{a}\right) + \frac{229}{40320} \frac{1}{a} \log\left(\epsilon\right)\right] . \label{21}
\end{eqnarray}
For Eq. (\ref{17a}), corresponding to $j = 0 $, we obtain
\begin{eqnarray}
E _{I,0} &=& \frac{\hbar c}{\pi} \left[- \frac{1}{24} \frac {\pi^{2} }{a} + \frac {1}{2} \frac{a}{\sigma ^{2}}\right]. \label{22}
\end{eqnarray}
So the energy of a scalar field considering a spherical configuration due the internal modes is
\begin{eqnarray}
E _{I} &=& E _{I_{partial}} + E _{I,0} \nonumber \\
&=& -\frac{\hbar c}{\pi a} 0.0017179275 + \frac{\hbar c}{\pi} \left[ \frac{1}{3}\frac{a}{\sigma^{2}} +2 \frac{a^{3}}{\sigma^{4}} \right.\nonumber\\
&-& \frac{17}{1920}\frac{1}{a}\log\left(\frac{\sigma}{a}\right) +\left. \frac{229}{40320} \frac{1}{a}\log\left(\epsilon\right)\right]\, . \label{23}
\end{eqnarray}

The expression (\ref{23}) shows in an undoubted way the need for a second regularized exponential function, $\exp(- \epsilon \nu) $, to makes possible a consistent mathematical handling of the divergences. Both divergences, the logarithm in (\ref{20a}) and the polynomial in (\ref{20b}), stem from the summation on $j$. This type of divergence already was observed in reference \cite{Brevik83} but only with the procedure established here this discard turns to be completely justified. 


\vspace*{2mm}

\subsection{External mode}

\indent

The contribution of the external modes we come by the Eq. (\ref{14b}). We proceed in an analogous way as the previous subsection. So
\begin{eqnarray}
\mathcal{E}_{O} =  E _{O} - E _{O} ^{(ref)} , \label{reg-outer}
\end{eqnarray}
where $E _{O} = E _{O,0} + E _{O,1} + E _{O,2} + E _{O,3} + E _{O,4} $ and
\begin{widetext}
\begin{eqnarray}
E _{O,0} &=& - \frac{\hbar c}{2 \pi}\Re \exp (-\epsilon \frac{1}{2} ) \exp (-i\varphi ) \int_{0}^{\infty } d\rho \exp (-i\sigma \rho \exp (-i \varphi) ) \left[ -\frac{1}{2} - \exp (-i \varphi)  a \rho \right], \label{24a} \\
E _{O,1} &=& \frac{\hbar c}{\pi a}\sum_{j=1}^{\infty }\nu ^{2}\int_{0}^{\infty}d\rho \left\{ \log\left[\mathcal{K} _{O} \left(\nu ,\rho\right)\right] -\sum_{k=1}^{4} \frac{\mathcal{U} _{(O,k)} \left(t\right)}{\nu ^k} \right\} \; ,\label{24b} \\
E _{E,2} &=& - \frac{\hbar c}{\pi}\Re \sum_{k=1} ^{4}\sum_{j=1}^{\infty }\nu ^{2-k}\exp (-\epsilon \nu ) \int_{0}^{\infty \exp (-i\varphi)} dz\exp (-i\sigma \nu z) z \frac{d}{dz}\mathcal{U}_{(O,k)} \left(t\right) \; ,\label{24c}\\
E _{O,3} &=&  \frac{\hbar c}{2 \pi}\Re \sum_{j=1}^{\infty }\nu ^{2} \exp (-\epsilon \nu ) \int_{0}^{\infty \exp (-i\varphi )}dz\exp (-i\sigma \nu z) \frac{a^{2}z^{2}}{1+a^{2}z^{2}} \; ,\label{24d}\\
E _{O,4} &=& \frac{\hbar c}{\pi}\Re \sum_{j=1}^{\infty }\nu ^{3} \exp (-\epsilon \nu ) \int_{0}^{\infty \exp (-i\varphi )} dz\exp (-i\sigma \nu z) \sqrt{1+a^{2}z^{2}} \; ,\label{24e}
\end{eqnarray}
\end{widetext}
with the following definitions \cite{Romeo96}
\begin{eqnarray}
\mathcal{K} _{O}\left(\nu ,\rho\right) &=& \sqrt{\frac{2\nu}{\pi} }\left( 1 + \rho ^{2}\right) ^{\frac{1}{4}} \exp\left(\nu \eta \right) K _{\nu} \left(\nu \rho \right), \label{25a} \\
\mathcal{U}_{(O,1)}\left(t\right) &=& - \mathcal{U}_{(I,1)} \left(t \right), \label{25b}
\end{eqnarray}
\begin{eqnarray}
\mathcal{U}_{(O,2)}\left(t\right) &=& \mathcal{U}_{(I,2)} \left(t \right) , \label{25c} \\
\mathcal{U}_{(O,3)}\left(t\right)&=& - \mathcal{U}_{(I,3)}\left(t \right), \label{25d} \\
\mathcal{U}_{(O,4)}\left(t\right)&=& \mathcal{U}_{(I,4)} \left(t \right), \label{25e}
\end{eqnarray}
where the $\mathcal{U} _{ \left(O, k \right) } $ are given by (\ref{18b}) to (\ref{18e}), respectively. The term (\ref{25b}) was numerically determined by \cite{Romeo96}
\begin{eqnarray}
E _{O,1} = - 0.00054 \frac{\hbar c}{\pi a} . \label{26}
\end{eqnarray}
The others contributions are calculated by following the analogous procedure detailed in the previous subsection.
\begin{eqnarray}
E _{O,2} &=& \frac{\hbar c}{\pi}\Re \left\{-\frac{5821}{56688} \frac{1}{a} - \frac{7}{24} \frac{a}{\sigma ^{2}} - \frac{11 }{192} \frac{1}{a}\log \left( \frac{\sigma}{a} \right) \right. \nonumber \\ &-& \left.\frac{229}{40320} \frac{1}{a} \log \left( \epsilon \right) - i \frac{7801}{86684} \frac{1}{a} \right\}, \label{27a}\\
E _{O,3} &=& \mathcal{E}_{I,3} \label{27b}\\
E _{O,4} &=& - \mathcal{E}_{I,4} \, . \label{27c}
\end{eqnarray}
After collect the terms (\ref{26}), (\ref{27a}), (\ref{27b}) and (\ref{27c}) we have
\begin{eqnarray}
E _{O _{partial}} &=& 0.01056399145 \frac{\hbar c}{\pi a} + \frac{\hbar c}{\pi} \left[\frac{1}{6}\frac{a}{\sigma^{2}} - 2 \frac{a ^{3}}{\sigma ^{4}}  \right. \nonumber\\
&+&\left.\frac{17}{1920} \frac{1}{a} \log\left( \frac{\sigma}{a}\right) - \frac{229}{40320} \frac{1}{a} \log \left( \epsilon \right) \right] . \label{28}
\end{eqnarray}
The contribution (\ref{25a}), related to $j = 0 $, when we repeat the calculation gives
\begin{eqnarray}
E _{O,0} &=& \frac{\hbar c}{\pi} \left[ - \frac{1}{2} \frac{a}{\sigma ^{2}} \right]. \label{29}
\end{eqnarray}
Gather together (\ref{28}) and (\ref{29}) we get the total contribution to the energy of a scalar field of a spherical configuration due the external modes
\begin{eqnarray}
E _{O} &=& E _{O_{partial}} + E _{O,0} \nonumber \\
&=& \frac{\hbar c}{\pi a} 0.01056399145 + \frac{\hbar c}{\pi} \left[ -\frac{1}{3} \frac{a}{\sigma ^{2}} - 2 \frac{a ^{3}}{\sigma ^{4}} \right. \nonumber\\
&+& \left. \frac{17}{1920} \frac{1}{a} \log \left( \frac{ \sigma}{a} \right) - \frac{229}{40320} \frac{1}{a} \log \left( \epsilon \right) \right] .  \label{30}
\end{eqnarray}
Our next task is determine the reference energy and take the regularizations as indicated by (\ref{reg-inner}) and (\ref{reg-outer}).


\subsection{The regularized results}

\indent

The last step is get the regularized results and to do it we need calculate the reference energies. Taking in account the contribution of $j=0 $ term we obtain
\begin{eqnarray}
E _{\pm} ^{(ref)} &=& \pm \frac{R}{\xi} \left( \frac{f (\epsilon)}{\sigma ^{2}} - (-1) ^{\phi/\pi} \frac{2}{\sigma ^{2}} \right), \label{34}
\end{eqnarray}
where the plus sign refers to $\mathcal{E} _{I} $ and the minus sign to $\mathcal{E} _{O} $ and
$f ( \epsilon) = \exp \left(- \frac{\epsilon}{2} \right)\left( 3 \exp (\epsilon ) - 1 \right) \left( \exp (\epsilon ) - 1 \right) ^{-2} $.

Now we can gather together the internal (\ref{23}) and external (\ref{30}) contributions, taking into account (\ref{34}), to obtain the Casimir effect for a scalar field due to the presence of a spherical shell with radius $a $
\begin{eqnarray}
\mathcal{E} (a) = \frac{\hbar c}{a} 0.002815789609 . \label{33}
\end{eqnarray}
This result is free of divergences since we get an exact cancelation for the terms which depend on the cut-off parameters. The Eq. (\ref{33}) is in agreement with that obtained by \cite{Romeo96}, through the zeta function method, and with that in \cite{Bender94}, which uses the Green function formalism and the dimensional analytical extension (in this reference the starting point is the expression for the force).

We can explicit the form and nature of each divergent term in (\ref{23}), (\ref{30}) and (\ref{34}) as a function of the geometrical proprieties of boundary if we rewrite the divergent part of those assuming a dimensionless parameter $\varepsilon = \left( \epsilon a \right) ^{\frac{357}{155}} $, with dim$(\epsilon ) = L ^{-1} $, so
\begin{eqnarray}
\mathcal{E} _{\pm} &=& \pm \hbar c \left[ \frac{3}{\pi ^{2}} V (a) \frac{1}{\sigma ^{4}} - \frac{17}{3840 \pi ^{2} } S (a) \kappa ^{3} \log \left( \sigma \varepsilon \right) \right. \nonumber \\ &-& \left. \frac{1}{ 3 \pi ^{2}} S (a) \kappa \frac{1}{\sigma ^{2}} - \frac{1}{4 \pi ^{2}} S \left( \frac{R}{\xi} \right)  \kappa \left( \frac{R}{\xi} \right) \frac{f (\varepsilon)}{\sigma ^{2}} \right. \nonumber \\ &-& \left. (-1) ^{\phi/\pi} S \left( \frac{R}{\xi} \right) \frac{2}{\sigma ^{2}} \right], \,\,\,\,\,\,\,\,\,\,\, \label{31}
\end{eqnarray}
where the plus sign refers to the index $I $ while the minus sign to the index $O $ and $\kappa = 1/a $ is the curvature. In (\ref{31}) $V (a) $ is a volume, $S (a) $ is an area and $\sigma $ and $\epsilon $ are cut-off parameters. As we can see the second and fourth terms in (\ref{31}) explain why it is required two regulators to get a well defined expression for the Casimir energy of the scalar field. This is the same case when we consider an electromagnetic field (see \cite{Miltao06}). The result (\ref{33}) is in agreement with \cite{Fulling03}, except for the divergence due to $\log (\epsilon ) $ and the relative self-energy of the spherical shell, $f (\varepsilon)/\sigma ^{2} $ and $2(-1) ^{\phi/\pi} /\sigma ^{2} $, which are not mentioned by that.


\section{Conclusions}

\indent

Our purpose in this work is to confirm that the prescription, in (\ref{12}), works when we assume other fields in the presence of a spherical shell. In fact, the approach has succeed in demonstrating the cancelation of all types of divergences appearing in the expression for the Casimir energy. Besides, this calculation presented at this work founded in shows a desired agreement with the results in the literature, something that has been proved for the electromagnetic field \cite{Miltao06,Miltao08}. Furthermore, as it has been mentioned by \cite{MiltonK03,MiltonK04,MiltonK05} a better understanding of the quantum field theory necessarily involves the need to understand this infinities.

%
The authors wish to thank Dr. Ludmila Oliveira H. Cavalcante (DEDU-UEFS) for valuable help with English revision.


\onecolumngrid

\end{document}